\documentclass[a4paper]{jpconf} \usepackage{graphicx}  
\usepackage{float}
\RequirePackage{lineno} 

\def\Journal#1#2#3#4{{#1} {\bf #2}, #3 (#4)}

\def\NIMA{{Nucl. Instr. Meth.} A}

\def\NPA{{Nucl. Phys.} A}
\def\PLB{{Phys. Lett.}  B}
\def\PRL{{Phys. Rev. Lett.}}
\def\PRC{{Phys. Rev.} C}
\def\PRD{{Phys. Rev.} D}

\def\EPJ{{Eur. Phys. J.} C}

\begin{document}

\title{Dielectron production in 200 GeV p+p and Au+Au collisions at STAR}

\author{Yi Guo for the STAR collaboration}

\address{}

\ead{yiguo@rcf.rhic.bnl.gov}

\begin{abstract} 
  Leptons do not interact strongly with the hot dense medium created in
  relativistic heavy ion collisions. They can escape the interaction region
  undistorted and thus carry direct information about the space-time evolution
  of the expanding system. In the low mass region (LMR, $0.3<M_{ee}<1.1$
  GeV/$c^{2}$), dielectron mass spectra can provide the in-medium vector meson
  properties, while in the intermediate mass region (IMR, $1.1<M_{ee}<3$
  GeV/$c^{2}$), the slope of dielectron transverse mass spectra is expected to
  have connection with the QGP temperature.

  In this paper, we present the centrality and $p_{T}$ dependence of the
  dielectron mass spectra measured in STAR experiment at RHIC. The data sets
  used in the analysis include large statistics samples collected during years
  2010 for 200 GeV Au+Au collisions and 2012 for 200 GeV p+p collisions. In
  order to extract underlying physics, we will compare our results with model
  calculations.
\end{abstract}

\section{Introduction}
Dileptons are clean and penetrating probes for the hot and dense nuclear matter
created by the high energy nuclear collisions because they do not suffer from
strong interactions. They can be produced during all stages of a heavy ion
collision, and their sources are expected to have different contributions to
dileption invariant mass spectra. Therefore, a systematic measurement of the
dilepton pair distribution can reveal the properties of medium created by high
energy nuclear collisions.

Dilepton measurements have been pursued for decades in heavy ion collisions
\cite{DLS}-\cite{PHENIX}. The CERES measurement of $e^{+}e^{-}$ mass spectra
showed a clear enhancement in the mass region below $\scriptsize{\sim}$0.7
GeV/$c^{2}$ compared to the known hadronic sources \cite{CERES}. High precision
data from NA60 suggested that this enhancement is consistent with in-medium
broadening of the $\rho$ spectral function instead of a dropping of its pole
mass hypothesis \cite{NA60}, \cite{Rapp4SPS}-\cite{PHSD4SPS}. In addition, slope
parameters of dimuon transverse mass spectra showed a sudden drop above the
$\phi$ mass after removing the correlated charm contributions.  This is argued
to be an indication of thermal dilepton from partonic source by the NA60
collaboration \cite{NA602}. At RHIC energy, the result from PHENIX showed a
significant enhancement in mass region 0.3$\scriptsize{\sim}$0.76 GeV/$c^{2}$
\cite{PHENIX}.  However, the huge enhancement could not be reproduced by
those model calculations which successfully explained SPS data
\cite{Rapp4SPS}-\cite{PHSD4SPS}.

In this paper, we will present the newest STAR results on dielectron production
from 200 GeV p+p and Au+Au collisions, and compare the Au+Au results with model
calculations. 

\section{Analysis} 
Data used in the analysis are obtained from 200 GeV p+p and Au+Au
collisions, which were collected by the Solenoidal Tracker At RHIC (STAR)
detector \cite{STARdet} in year 2012 and year 2010, respectively. The main
subsystems used in the analysis are Time Projection Chamber(TPC) \cite{STARTPC}
and the Time of Flight (TOF) \cite{STARTOF}.

In addition of track reconstruction and momentum information, TPC also
provides identification capabilities for charged particles by their ionization
energy loss ({\it dE/dx}) in the TPC gas. With the fully installed TOF system,
the PID capabilities are greatly improved, especially in low $p_{T}$ region.
With the combination of {\it dE/dx} from TPC and velocity ($\beta$) from TOF,
the electron purity is $\scriptsize{\sim}$98\% in p+p collisions,
$\scriptsize{\sim}$95\% in Au+Au minimum bias collisions and
$\scriptsize{\sim}$93\% in Au+Au central collisions.

\begin{figure}[H]
  \begin{minipage}{18pc}
    \begin{center}
      \includegraphics[width=16pc]{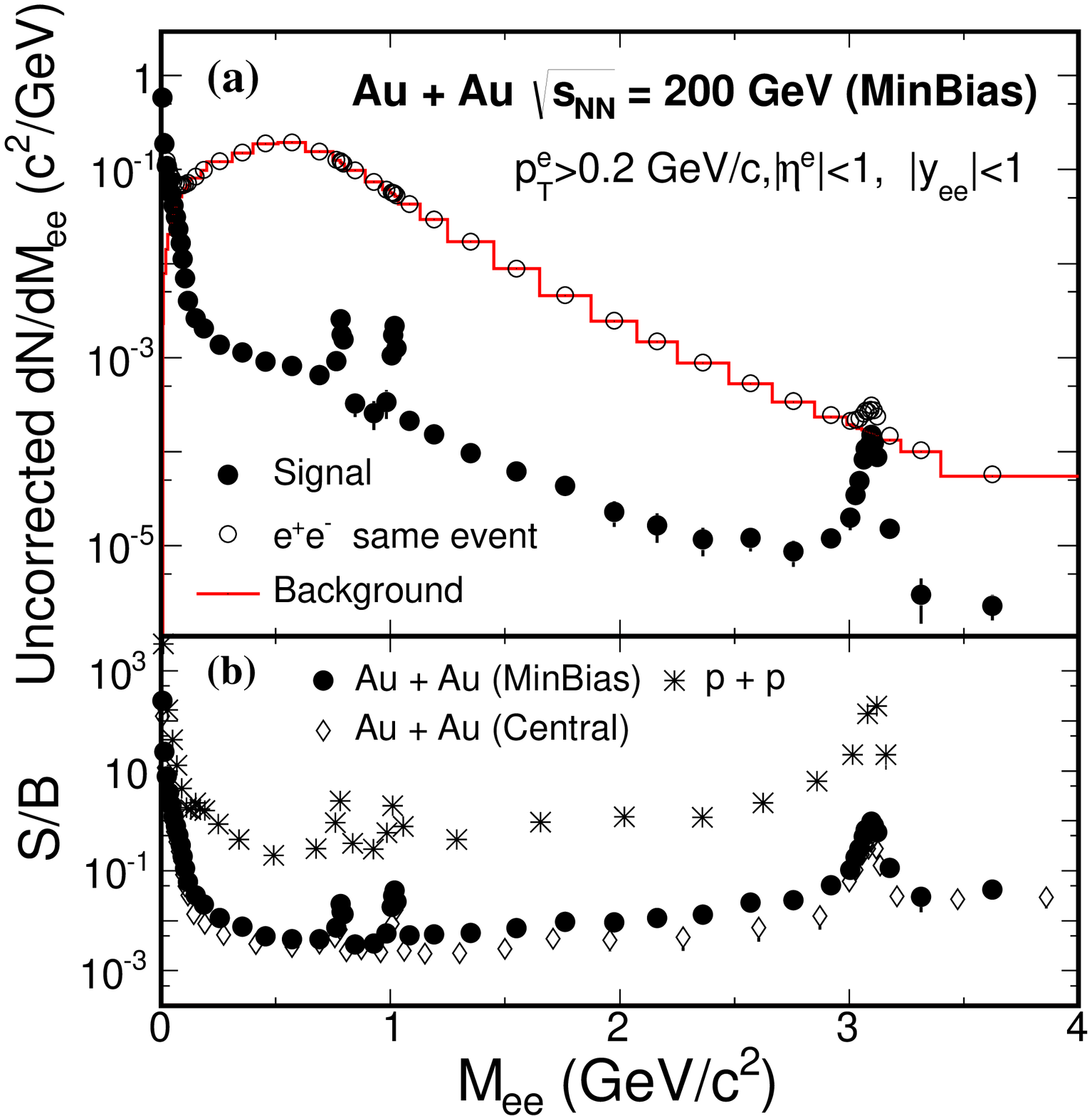}
    \end{center}
  \end{minipage}
  \begin{minipage}{18pc}
    \begin{center}
      \includegraphics[width=3in]{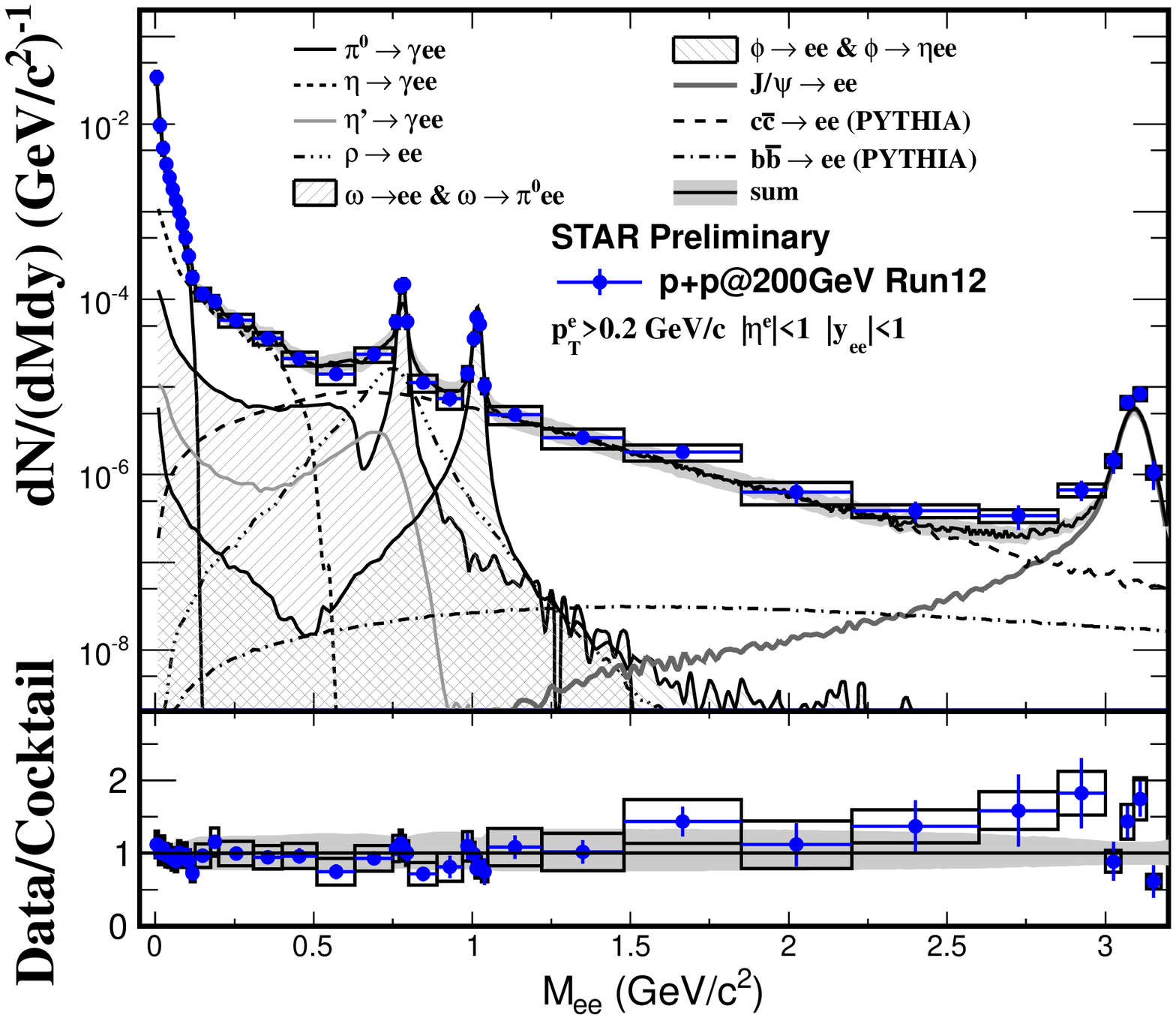}
    \end{center}
  \end{minipage}
    \caption{
      Left panel (a): $e^{+}e^{-}$ raw invariant mass distribution (open
      circles), the reconstructed backgrounds (red histogram) and the signal
      (solid dots) in 200 GeV Au+Au minimum bias collisions \cite{PRLaxiv}. Left
      panel (b): The signal to background ratio in p+p and Au+Au collision at
      $\sqrt{s_{NN}}$ = 200 GeV \cite{PRLaxiv}.  Right panel: invariant mass
      spectra from $\sqrt{s}$ = 200 GeV p+p collisions taken from year 2012. The
      black open box represents systematic error from data while the grey band
      depicts systematic uncertainty of cocktail.
    }
  \label{fig:bg}
\end{figure}

Two methods were used in this analysis to reconstruct the background. In LMR,
due to the correlated background, e.g. cross pair and jet contribution, we used
the like-sign method background with an acceptance correction. The acceptance
correction is to account for the slight difference in the acceptance between
like-sign and unlike-sign pairs \cite{STARPP}. While in IMR, the mix-event
technique was used to achieve better statistics. The difference between these
two methods were taken as systematic uncertainty. Figure~\ref{fig:bg} left panel
(a) shows the raw dielectron mass distribution, reconstructed background and
signal after background subtraction in 200 GeV Au+Au minimum bias collisions.
The signal-to-background ratios from p+p minimum bias collisions, Au+Au minimum
bias and central collisions are shown in Fig.~\ref{fig:bg} left panel (b). In
this analysis, we also subtracted the contribution from the photon conversion by
the method used in Ref. \cite{PHENIXPhoton}.

The dielectron continuum results in this paper were obtained in STAR acceptance
($p_{T}^{e} > 0.2$ GeV/$c$, $|\eta^{e}|<1.0$, $|y^{ee}|<1.0$) and corrected for
the efficiency.

\section{Results}
Figure~\ref{fig:bg} right panel shows the dielectron invariant mass spectra from
200 GeV p+p collisions taken in year 2012. The cocktail is taken from the STAR
published result \cite{STARPP}, and the charm cross section is updated to
$797\pm210$(stat.)$^{+208}_{-295}$(sys.)$\mu$b with respect to the newest
published result from STAR \cite{STARCharm}.  The cocktail simulation can
reproduce the new preliminary result very well. With a full TOF coverage and
more data taken, year 2012 result has greatly improved statistics which is
$\scriptsize{\sim}$7 times more than STAR published result \cite{STARPP}. The
large statistics new results at p+p 200 GeV provide a better baseline for Au+Au
collisions.

The $\sqrt{s_{NN}}$ = 200 GeV Au+Au results taken in year 2010 has been accepted
by Phys.Rev.Lett \cite{PRLaxiv}. In this paper, we will briefly review the key
content of the submitted paper.  

\begin{figure}
  \begin{minipage}[b] {1.0\textwidth}
    \begin{center}
      \includegraphics[width=2.5in]{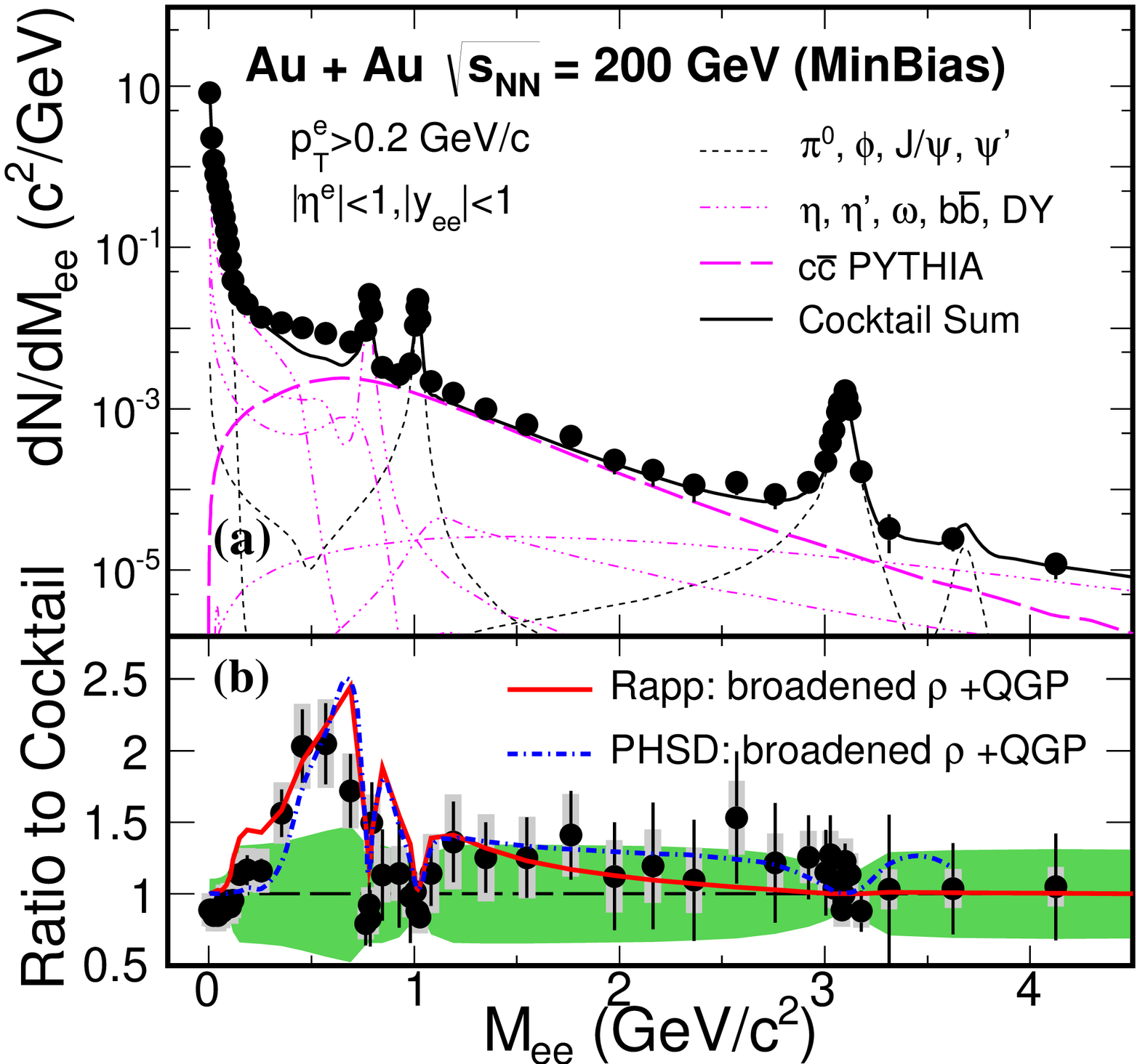}
    \end{center}
  \end{minipage}
  \begin{minipage}[b] {1.0\textwidth}
    \begin{center}
      \includegraphics[width=2.5in]{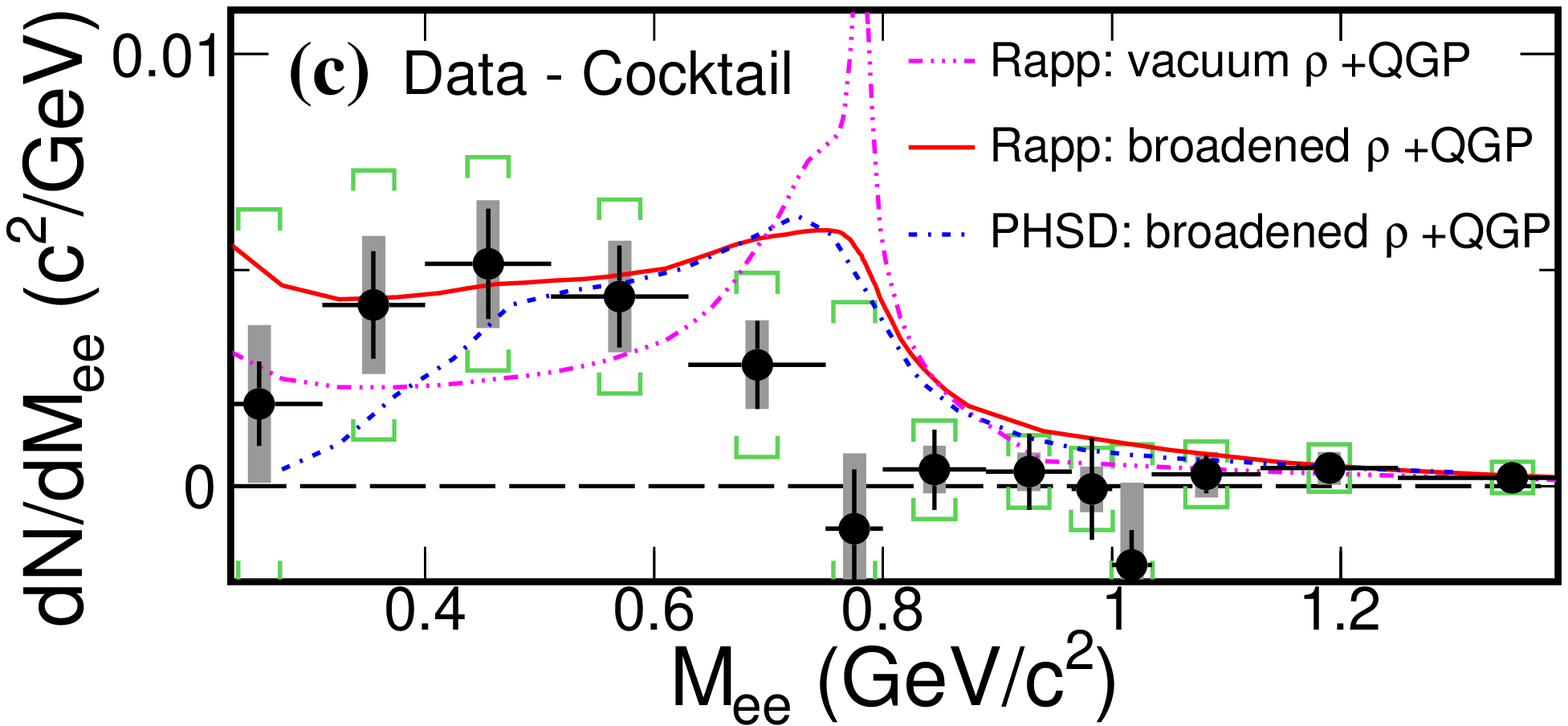}
    \end{center}
  \end{minipage}
  \caption{
    (a) Invariant mass spectra, (b) ratio of data to cocktail and (c) excess
    spectra in LMR from $\sqrt{s_{NN}}$ = 200 GeV Au+Au minimum bias collisions
    \cite{PRLaxiv} . Two model calculations are also included. In panel (b), the
    grey box represents the systematic uncertainty from data while the light
    green band shows the systematic uncertainty from cocktail. In panel (c),
    green brackets depict the total systematic uncertainties including those
    from cocktail.
  }
  \label{fig:AuAu200MB}
\end{figure}

In LMR, an enhancement of 1.77$\pm$0.11(stat.)$\pm$0.24(sys.)$\pm$0.41(cocktail)
is observed with respect to the cocktail without $\rho$, in the mass region
0.3$\scriptsize{\sim}$0.76 GeV/$c^{2}$ in minimum bias collision
(Fig.~\ref{fig:AuAu200MB} (b)). In addition, two model calculations \cite{Rapp,
PHSD} are included to compare with our data (Fig.~\ref{fig:AuAu200MB} (b), (c)):
Model I by Rapp {\it et al.} is an effective many-body calculation
\cite{Rapp4SPS, Rapp}; Model II by Linnyk {\it et al.} is a microscopic
transport model, Parton-Hadron String Dynamics (PHSD) \cite{PHSD4SPS,PHSD}. Both
models involve in-medium broadened $\rho$ spectral function hypothesis and can
successfully reproduce the NA60 results.  The models, however, failed to
reproduce the enhancement in central collisions reported by the PHENIX
experiment \cite{PHENIX, PHSD}.  In the mass region below 1 GeV/$c^{2}$, both
models describe our data reasonably well within uncertainties.  Our measurements
disfavor a pure vacuum $\rho$ mass distribution for the excess dielectron
($\chi^{2}/NDF = 25/8$ in 0.3-1GeV/$c^{2}$, Fig.~\ref{fig:AuAu200MB} (c)).

\begin{figure}[H]
  \begin{center}
    \includegraphics[width=4.5in]{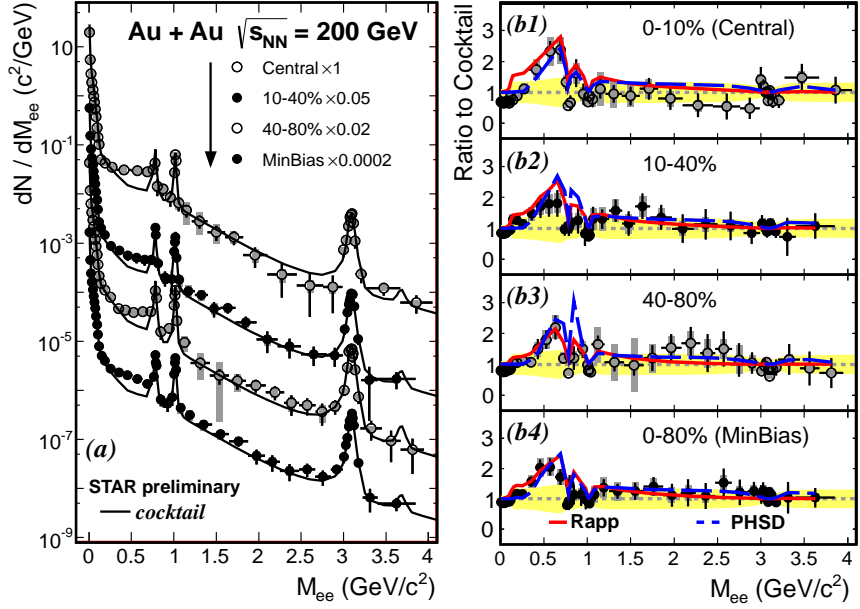}
  \end{center}
  \caption{Left panel shows dielectron invariant mass spectra in different
    centralities. The solid curves represent the hadronic cocktail. The charm
    contribution is calculated by PYTHIA and scaled by $N_{\rm{bin}}$. Right
    panel shows the ratio of data to cocktail in different centralities. The
  yellow band represents the systematic uncertainty of cocktail.}
  \label{fig:cen}
\end{figure}

\begin{figure}[H]
  \begin{center}
    \includegraphics[width=4.5in]{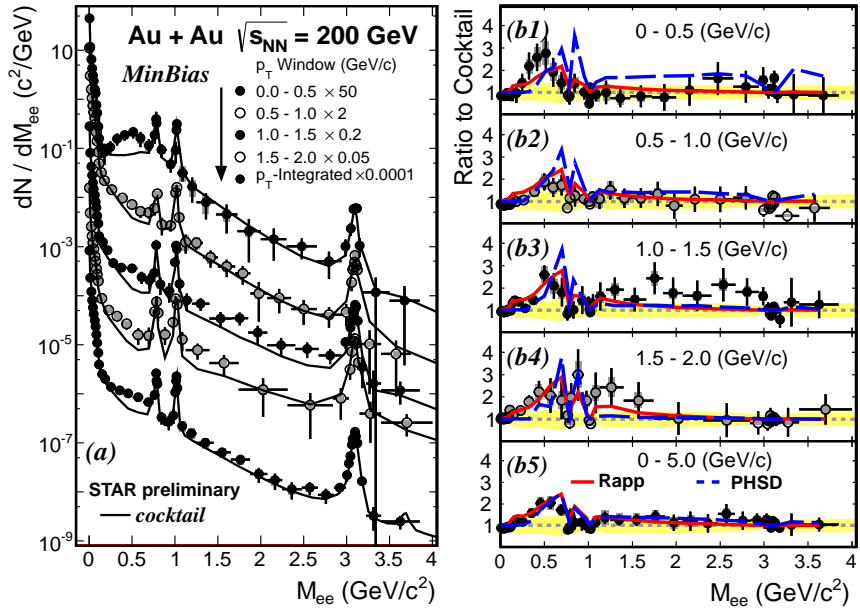}
  \end{center}
  \caption{Left panel shows dielectron invariant spectra in different
    $p^{ee}_{T}$ ranges. The solid curves represent the hadronic cocktail.
    Right panel shows the ratio of data to cocktail in different $p^{ee}_{T}$
    ranges. The yellow band represents the systematic uncertainty of cocktail.}
\label{fig:pT}
\end{figure}

Figures~\ref{fig:cen} and~\ref{fig:pT} show the dielectron spectra
measured in various centrality bins and $p_{T}$ ranges, respectively. The ratios
between data and cocktail are shown in the right panels.  Model calculations are
also included as a comparison. Both models are able to describe the LMR excess
in all $p_{T}$ and centrality bins within uncertainty.

\begin{figure}
\begin{center}
    \includegraphics[width=3in]{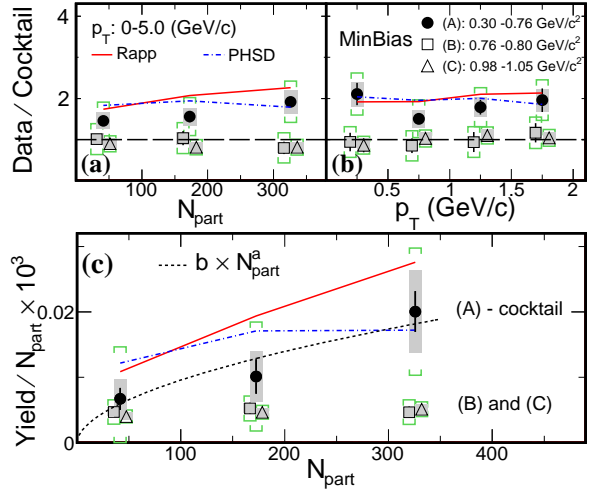}
\end{center}
\caption{
  Panel (a) and (b) shows the integrated dielectron yields in mass regions of
  0.3-0.76($\rho$-like), 0.76-0.80($\omega$-like) and 0.98-1.05($\phi$-like)
  GeV/$c^{2}$ as a function of centrality and $p_{T}^{ee}$. Panel (c) shows the
  yields scaled by $N_{\rm{part}}$ for the $\rho$-like with cocktail
  subtraction, and the $\omega$-like and $\phi$-like without cocktail
  subtraction \cite{PRLaxiv}. The dashed curve is a power-law fit to the
  yield/$N_{\rm{part}}$ for the $\rho$-like region subtracted by cocktail.
  Systematic uncertainties from data are shown as grey boxes, while the green
  brackets represent the total systematic uncertainties including the cocktail
  contribution. The $\omega$-like and $\phi$-like data points are slightly
  displaced horizontally for clarity.
}
\label{fig:IntY}
\end{figure}

We also report the ratios of data to cocktail within STAR acceptance in three
different mass regions: 0.3-0.76 ($\rho$-like), 0.76-0.8($\omega$-like) and
0.98-1.05($\phi$-like) GeV/$c^{2}$ as a function of centrality
Fig.~\ref{fig:IntY} (a) and dielectron $p_{T}$ Fig.~\ref{fig:IntY} (b). The
hadronic cocktail can reproduce the dielectron yield in the $\omega$-like and
$\phi$-like regions. In the $\rho$-like region, a significant excess is observed
and the ratio of data to cocktail shows a weak dependence on $N_{\rm{part}}$ and
dielectron $p_{T}$. Figure ~\ref{fig:IntY} (c) shows the yields in the
$\rho$-like region subtracted by cocktail, and the $\omega$-like and $\phi$-like
regions without cocktail subtraction. Dielectron yields in the $\omega$-like and
$\phi$-like regions show a $N_{\rm{part}}$ scaling. The dashed curve is a power
fit ($\propto N_{\rm{part}}^{a}$) to the excess yield/$N_{\rm{part}}$ in the
$\rho$-like region, and the fit result shows $a = 0.54\pm0.18$
(stat.+uncorrelated sys.), indicating the dielectron excess yields in the
$\rho$-like region are sensitive to the QCD medium dynamics, as expected by the
theoretical calculations \cite{Rapp,RhoOmega}.

\begin{figure}[H]
  \begin{center}
    \includegraphics[width=3in]{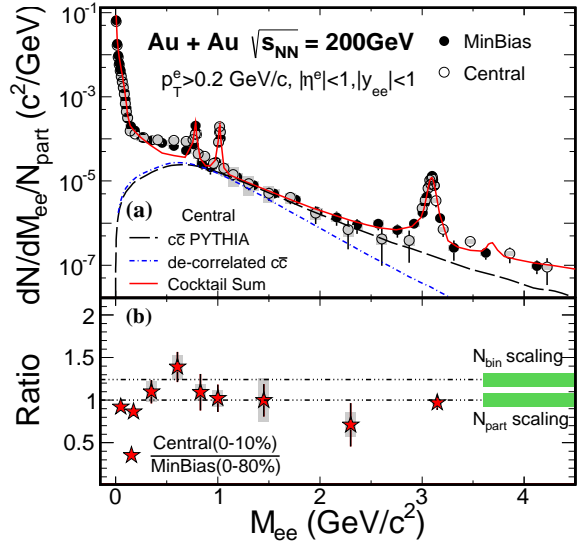}
  \end{center}
  \caption{
    (a) Dielectron invariant mass spectra from minimum bias (0-80\%) and central
    (0-10\%) collisions \cite{PRLaxiv}. The spectra are scaled by the number of
    participant nucleons ($N_{\rm{part}}$). The solid line represents the
    hadronic cocktail for central collisions. The dashed line depicts the
    correlated charm from PYTHIA, while the dot-dashed line assumes a fully
    randomized azimuthal correlation between charm pairs and the $p_{T}$
    suppression factor on single electron spectrum from RHIC is also included
    \cite{PHENIXNPE}. (b) The ratio of $N_{\rm{part}}$ scaled dielectron yields
    between the central and minimum bias collisions.  Systematic uncertainties
    are shown as the grey bands.
  }
  \label{fig:Rcp}
\end{figure}

In Fig.~\ref{fig:Rcp}, we overlay the dielectron mass spectra from minimum bias
and most central (0-10\%) collisions. The spectra are scaled by the number of
participant nucleons ($N_{\rm{part}})$. The ratio in the bottom panel starts
from unity in the $\pi^{0}$ and $\eta$ mass region and begins to increase in
mass region 0.5-1 GeV/$c^{2}$ towards the $N_{\rm{bin}}$. This is due to the
fact that correlated charm contribution starts to dominate in this mass region
and the charm quark production at RHIC energy is expected to follow the
$N_{\rm{bin}}$ scaling. The hadronic medium also has a significant contribution
in this mass region and is expected to increase faster than $N_{\rm{part}}$
\cite{Rapp,RhoOmega}. In the IMR region, the ratio shows a moderate deviation
from the $N_{\rm{bin}}$ scaling (1.8$\sigma$ deviation for the data point at
1.8-2.8 GeV/$c^{2}$). The difference in mass region 1-3 GeV/$c^{2}$ indicates a
potential de-correlating effect on charm pairs while traversing the QCD medium
or other contribution from medium (e.g thermal radiation).

\section*{References} 
\begin{thebibliography}{999} 
    \bibitem{DLS} R.J. Porter {\it et al.},
    \Journal{\PRL}{79}{1229}{1997}.

    \bibitem{HADES} G. Adamova {\it et al.},
    \Journal{\PRL}{98}{052302}{2007}; \Journal{\PRC}{84}{014902}{2011}.

    \bibitem{HELIOS} A.L.S Angelis {\it et al.}, 
    \Journal{\EPJ}{13}{433}{2000}.

    \bibitem{CERES} G. Agakichiev {\it et al.},
    \Journal{\PRL}{91}{042301}{2003}; \Journal{\EPJ}{41}{475}{2005}.

    \bibitem{NA60} R. Arnaldi {\it et al.},
    \Journal{\PRL}{96}{162302}{2006}; \Journal{\PRL}{100}{022302}{2008};

    \bibitem{NA602} R. Arnaldi {\it et al.},
    \Journal{\EPJ}{59}{607}{2009}.

    \bibitem{PHENIX} A. Adare {\it et al.},
    \Journal{\PRC}{81}{034911}{2010}.

    \bibitem{Rapp4SPS} H. van Hees and R. Rapp, 
    \Journal{\PRL}{97}{102301}{2006}; 
    H. van Hees and R. Rapp, \Journal{\NPA}{806}{339}{2008}.

    \bibitem{Renk4SPS} J. Ruppert, C. Gale, T. Renk, P. Lichard and J. Kapusta, 
    \Journal{\PRL}{100}{162301}{2008}; 
    T. Renk and J. Ruppert, \Journal{\PRC}{77}{024907}{2008}.

    \bibitem{Dusling4SPS} K. Dusling, D. Teaney and I. Zahed, 
    \Journal{\PRC}{75}{024908}{2007}.

    \bibitem{PHSD4SPS} O. Linnyk {\it et al.}, 
    \Journal{\PRC}{84}{054917}{2011}.

    \bibitem{STARdet} Special Issue on RHIC and Its Detectors, edited by
    M. Harrison, T. Ludlam, and S. Ozaki,
    \Journal{\NIMA}{499}{No. 2-3}{2003}.

    \bibitem{STARTPC} M. Anderson, {\it et al.},
    \Journal{\NIMA}{499}{659}{2003}.

    \bibitem{STARTOF} W.J. Llope (for the STAR Collaboration), 
    \Journal{\NIMA}{661}{S110}{2012}.

    \bibitem{STARPP} L. Adamczyk {\it et al.}, 
    \Journal{\PRC}{86}{024906}{2012}.

    \bibitem{PHENIXPhoton} A. Adare {\it et al.}, 
    \Journal{\PRC}{81}{034911}{2010}.

    \bibitem{STARCharm} L. Adamczyk {\it et al.}, 
    \Journal{\PRD}{86}{072013}{2012}.

    \bibitem{PRLaxiv} L. Adamczyk {\it et al.}, arXiv:1312.7397.

    \bibitem{Rapp} R. Rapp, PoS CPOD2013, 008 (2013); R. Rapp, private communications.

    \bibitem{PHSD} O. Linnyk {\it et al.}, 
    \Journal{\PRC}{85}{024910}{2012}; O. Linnyk, private communications.

    \bibitem{RhoOmega} U. Heinz and K.S. Lee, 
    \Journal{\PLB}{259}{162}{1991}.

    \bibitem{PHENIXNPE} A. Adare {\it et al.}, 
    \Journal{\PRL}{98}{172301}{2007}. 

\end{thebibliography}
\end{document}